\documentclass[12pt,a4paper]{article}
\usepackage{epsfig,graphics,subfigure} 
\usepackage{rotating}
\usepackage{latexsym}
\usepackage{nature}
\usepackage{tiny_headings}
\newcommand{\QWERTY}{{\sc qwerty}}
\newcommand{\labelstyle}[1]{\textsf{#1}}
\begin{document}
\title{Fast Hands-free Writing by Gaze Direction}
\author{}
\date{}
\maketitle
\begin{raggedleft}\raisebox{1.7in}[0in]{{\em Nature\/} {\bf 418}, p.\ 838 (22nd August 2002). {{\tt{www.nature.com}}}}\\[-0.9in]
\end{raggedleft}

{
 We describe a method for text entry based on
 inverse arithmetic coding
 that relies on  gaze direction
 and which is  faster and more accurate than  using an on-screen keyboard.
 These benefits are derived from  two innovations: the writing task is matched
 to the capabilities of the eye, and a language
 model is used to make predictable words and phrases easier to write.
}

 For people who cannot use a standard keyboard or mouse, the direction
 of gaze is one of the few ways of conveying information to a
 computer. Many systems for gaze-controlled text entry provide an
 on-screen keyboard whose buttons are `pressed' by staring at
 them. But eyes did not evolve to push buttons, and this method of
 writing is exhausting.

 Moreover, on-screen keyboards are 
 inefficient because typical text has considerable
 redundancy.\cite{Shannon48}  Although a partial solution to this defect is to
 include word-completion buttons as alternative buttons alongside the
 keyboard,  a language model's predictions can be better
 integrated into the writing process.
 By inverting an efficient method for text {\em compression} --
 arithmetic coding\cite{arith_coding} -- we have created an efficient
 method for text {\em entry}, one that is also well matched to the
 eye's natural talent for search and navigation.

 One way to write a
 piece of text is to go into the library that contains {\em all
 possible books},\cite{babel} and find the book that contains exactly
 that text.
 Writing thus becomes a navigational task.  In
 our idealized library, the `books' are arranged alphabetically on one
 enormous shelf.  As soon as the user looks at a part of the shelf,
 the view zooms in continuously on the point of gaze.  To write a
 message that begins `{\tt{hello}}', one first steers towards the
 section of the shelf marked {\tt{h}}, where all the books beginning
 with {\tt{h}} are found. Within this section are sections for books
 beginning {\tt{ha}}, {\tt{hb}}, {\tt{hc}}, etc.; one enters the
 {\tt{he}} section, then the {\tt{hel}} section within it, and so
 forth.

 To make the writing process efficient we use a language model, which
 predicts the probability of each letter's occurring in a given
 context, to allocate the  shelf-space for each
 letter of the alphabet (figure \ref{fig.hello}a).
 When the language model's predictions are accurate, many successive
 characters can be selected by a single gesture.

 We previously evaluated this system, which we
 call {\sl Dasher}, with a mouse as the steering device.\cite{ward2000}
 Novices
 rapidly learned to write and an expert could write at 34 words per
 minute. All users made fewer errors than when using a
 standard \QWERTY\ keyboard.

 Figure \ref{fig.eye1}b shows an evaluation of Dasher driven by an
 eyetracker, compared with an on-screen keyboard.  After an hour of
 practice, Dasher users could write at up to 25 words per minute,
 whereas on-screen keyboard users could write at only 15 words per
 minute. Moreover, the error rate with the on-screen keyboard was about five
 times that of Dasher.

 Users of both systems reported the on-screen keyboard more stressful
 than Dasher, for two reasons.  First, they often felt uncertain
 whether an error had been made in the current word (the
 word-completion feature works only if no errors have been made);  an error
 can be spotted only by looking away from the keyboard. Second,
 a decision has to be made after `pressing' each character on whether
 to use word completion or
 to continue typing; looking in the word-completion area is a gamble
 as it is not guaranteed that the required word will be there;
 finding the right completion requires a switch to a new mental
 activity.
 By contrast, Dasher users can see simultaneously the last few
 characters they have written and the most probable options for the
 next few.   Furthermore, Dasher makes no distinction between word-completion and
 ordinary writing.

 Dasher works in most languages; the language model can be trained on
 example documents, and adapts to the user's language as she writes.
 It can also  be operated with other pointing devices such as a touch
 screen or rollerball. Dasher  is potentially an efficient,
 accurate, and fun writing system not only for disabled
 computer users but also for users of mobile computers.

\vfill
\raggedleft
David J. Ward and David J.C. MacKay
\\
 Cavendish Laboratory,  Cambridge, CB3 0HE
\\
{\tt
mackay@mrao.cam.ac.uk
}\\

\vfill

\bibliographystyle{nature}
\bibliography{bibs}

%
%

\newcommand{\thiswidth}{3in} 
\newcommand{\bigshotmouse}[2]{%
\setlength{\unitlength}{\thiswidth}
\begin{picture}(0.6,0.99)(0.02,0.01) 
\put(0,0){\psfig{figure=#1.ps,height=\thiswidth}}
\put(#2){{\psfig{figure=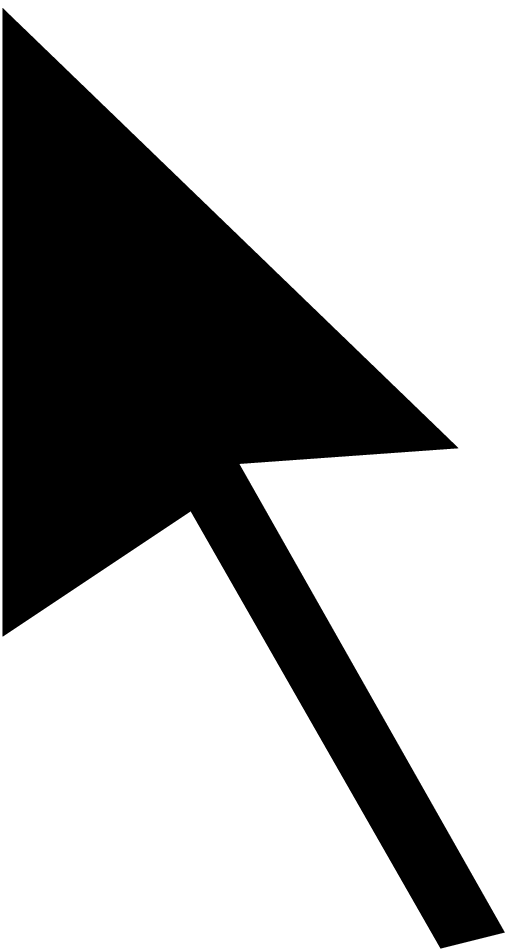,height=0.1207in}}}
\end{picture}
}
\begin{figure}[p]
\centering\small
\begin{tabular}{cc}
\raisebox{1.4in}{(a)}\begin{tabular}{c}
\bigshotmouse{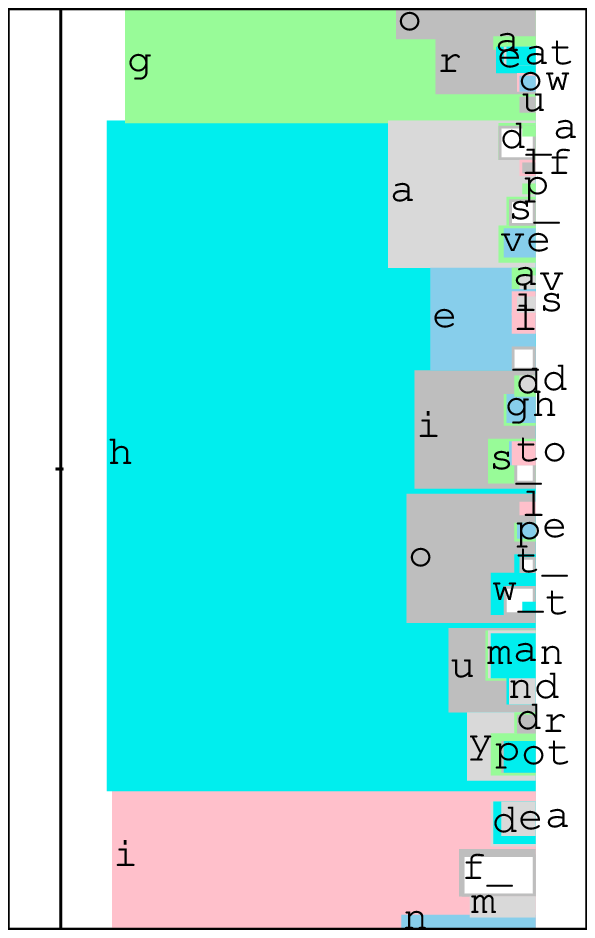}{0.5149975,0.620}
\end{tabular}
&
\raisebox{1.4in}{(b)}\hspace{-1mm}%
\begin{tabular}{@{}c@{}c@{\hspace*{1.52mm}}c@{\hspace*{1.2mm}}c} & & \labelstyle{ Dasher with eyetracker } & \labelstyle{On-screen keyboard } \\
{\begin{sideways}\labelstyle{ $\:\:\:\:\:$ Writing speed}\end{sideways} }& 
{\begin{sideways}\labelstyle{ $\:\:\:\:\:\:\:$ (words/min)}\end{sideways} }& 
 \includegraphics[width=3.7cm]{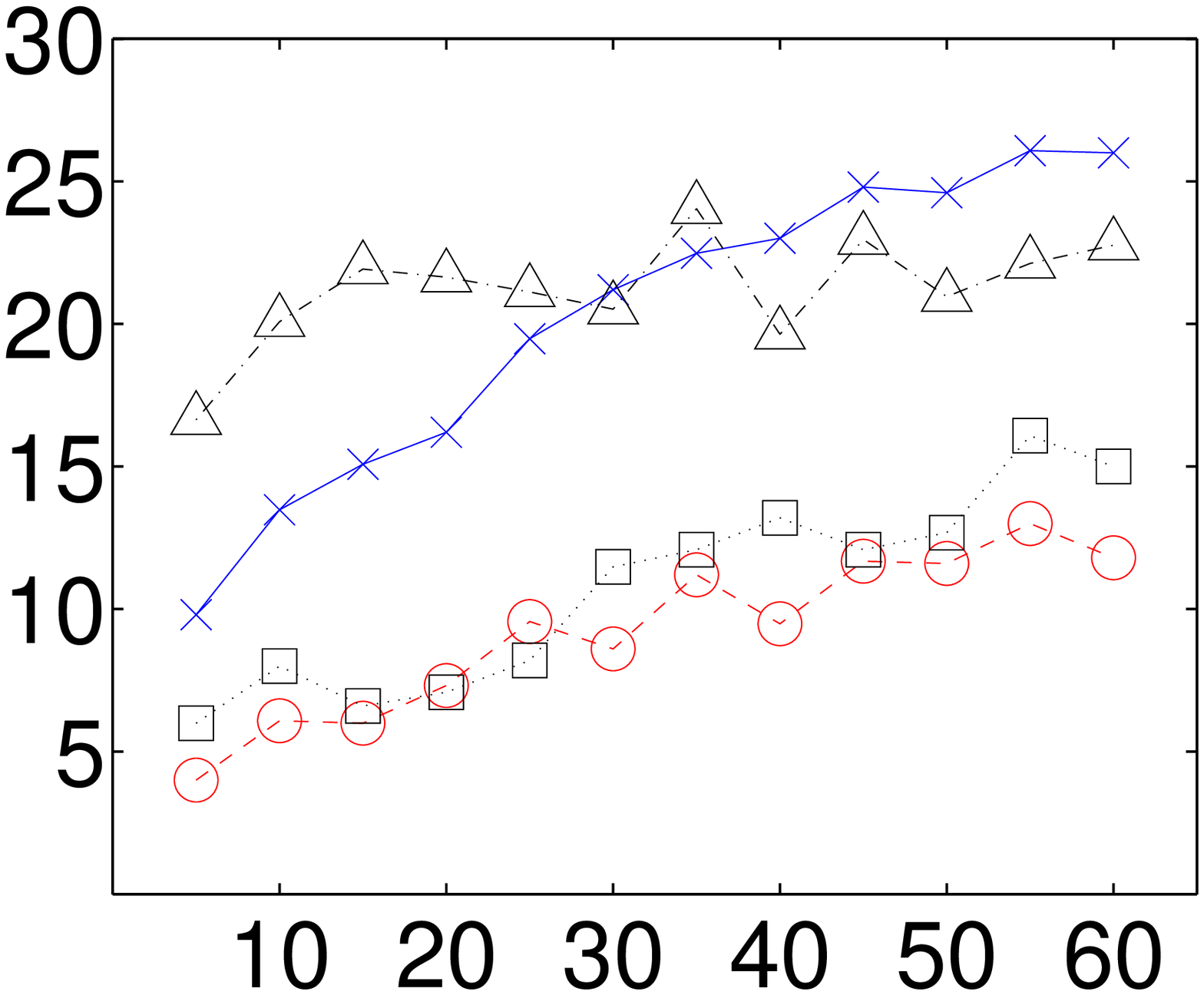} 
&
 \includegraphics[width=3.7cm]{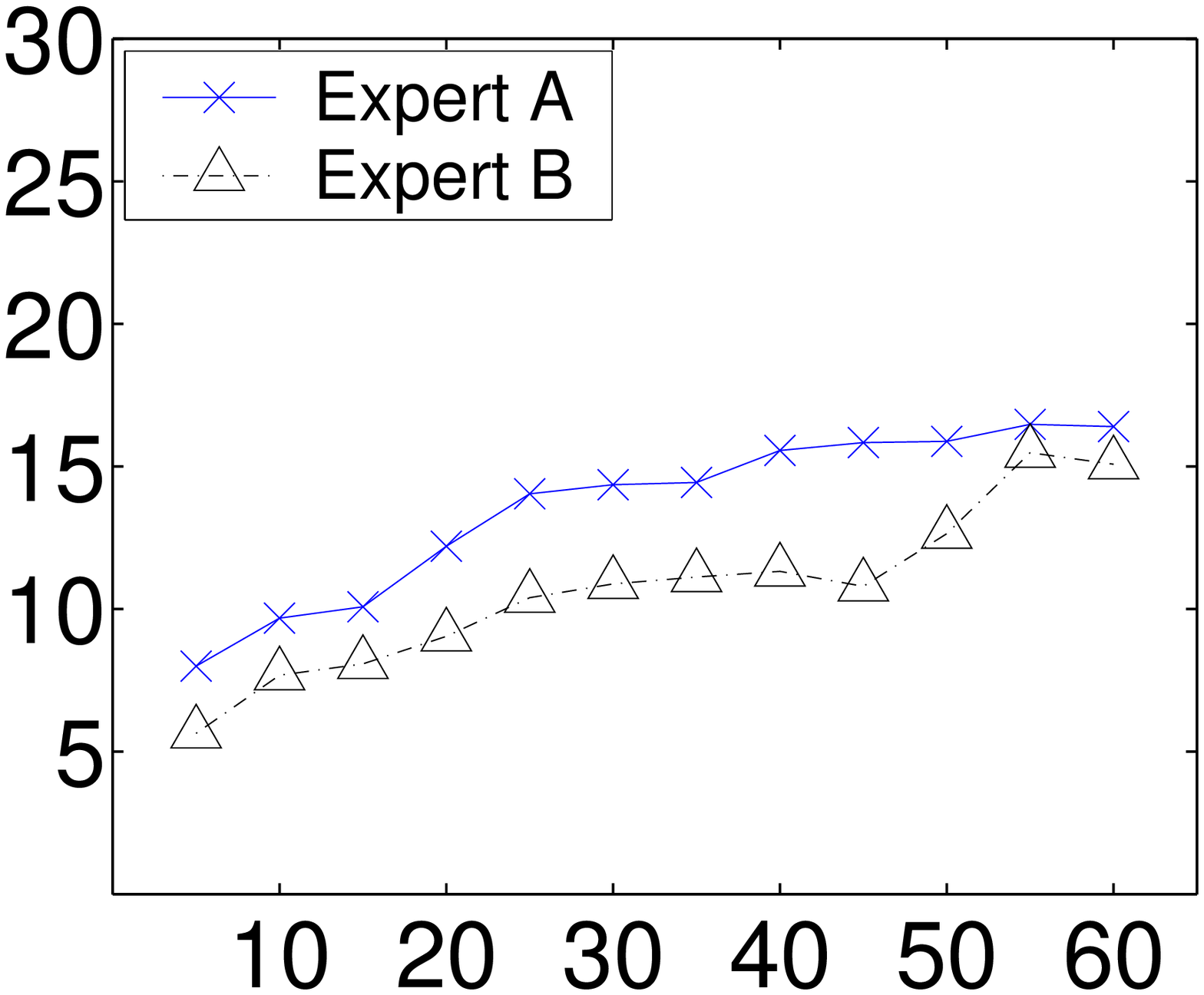} 
\\
\begin{sideways}\labelstyle{ $\:\:\:\:\:\:\:\:\:$ Percentage}\end{sideways}  &
\begin{sideways}\labelstyle{ $\:\:\:\:\:\:\:$ words wrong}\end{sideways}  &
  \includegraphics[width=3.7cm]{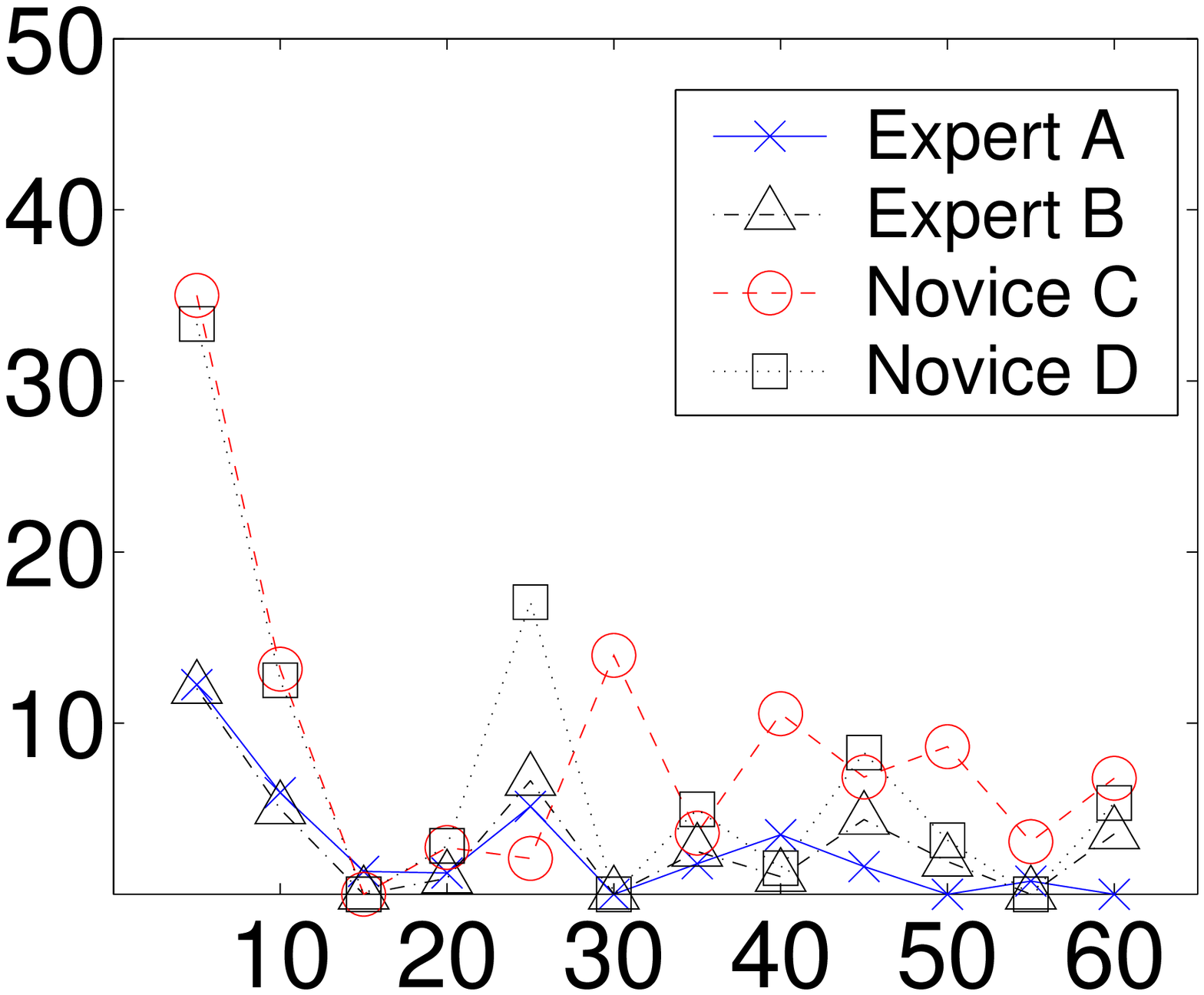} 
&  \includegraphics[width=3.7cm]{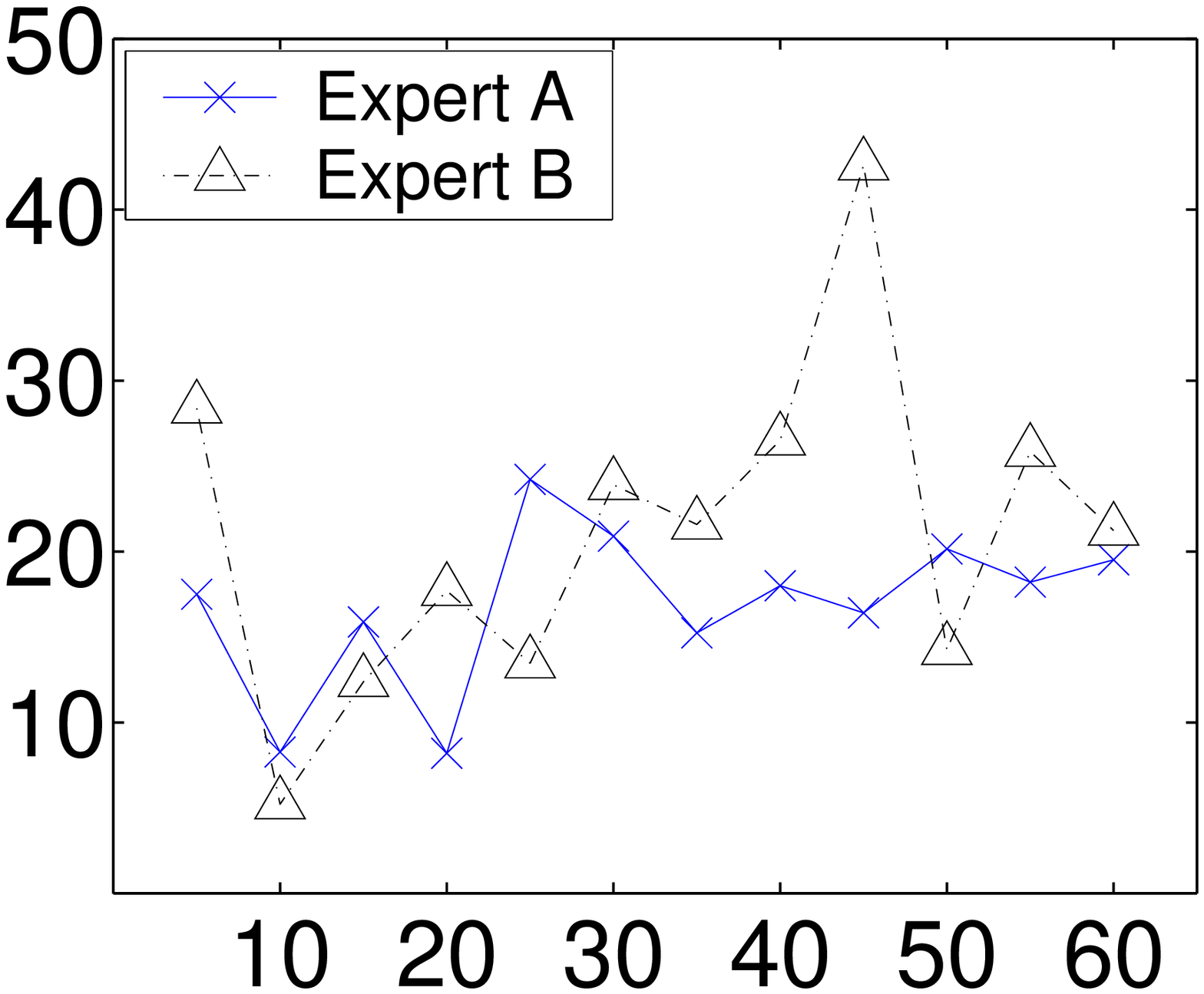}
\\
&
& \labelstyle{Time (mins)}
& \labelstyle{Time (mins)}
\\
\end{tabular}
\end{tabular}

\caption[a]{
 Hands-free text entry. 
 (a) Screenshot of Dasher when the user begins writing
 {\tt{hello}}. The shelf of the alphabetical `library' is displayed
 vertically. The space character, `{{\raisebox{-1mm}{$-$}}}', is
 included in the alphabet after {\tt{z}}. Here, the user has zoomed in
 on the portion of the shelf containing messages beginning with
 {\tt{g}}, {\tt{h}}, and {\tt{i}}.  Following the letter {\tt{h}}, the
 language model makes the letters {\tt{a}}, {\tt{e}}, {\tt{i}},
 {\tt{o}}, {\tt{u}}, and {\tt{y}} easier to write by giving them more
 space.  Common words such as {\tt{had}} and {\tt{have}} are visible.

 The arrow indicates the gaze of the user; its vertical coordinate
 controls the zooming-in point and its horizontal
 coordinate controls the rate of zooming; looking to the left makes
 the view zoom out, allowing  recent errors to be corrected.

 (b) Comparison of writing speeds and error rates
 for two methods of gaze-driven text entry.
 Left, Dasher\cite{dasher164} with eye-tracker, as
 recorded for two expert users of the system (crosses, triangles)
 and two novices (circles, squares); right,   on-screen
 keyboard, used by two experts on the \QWERTY\ keyboard.
 The eyetracking system was EyeTech's Quick
 Glance eyetracker.
 Each user took
 dictation from Jane Austen's {\em Emma\/} in five-minute sessions.
 The language model (PPMD5) predicts the
%
%
 next character given the previous five
 characters;\cite{cleary84,teahan95probability} it was trained on
 passages from {\em Emma\/} not included in the dictation.
 Right panels, the two experts took dictation using
 the same eyetracker to control the WiViK on-screen
 keyboard (a standard \QWERTY\ keyboard)
 with the word-completion buttons enabled. 
}
\label{fig.hello}
\label{fig.eye1}
\end{figure}
\end{document}